\documentclass[10pt]{article}
\usepackage{latexsym,graphicx}
\newcommand{\be}{\begin{equation}}
\newcommand{\ee}{\end{equation}}
\def\n{\noindent}
\catcode `\@=11
\catcode `\@=12
\begin{document}
\begin{center}
{\bf {BULK VISCOUS COSMOLOGICAL MODELS IN GENERAL RELATIVITY}} \\
\vspace{5mm}
\normalsize{ANIRUDH PRADHAN $^{\ast}$ \footnote{Corresponding Author}, VANDANA RAI $^{\dag}$ and 
R. S. SINGH {$^{\ddag}$}  } \\
\vspace{10mm}
\normalsize{$^{\ast}$\it{Department of Mathematics, Hindu Post-graduate College,
Zamania-232 331, Ghazipur, India \\
E-mail : pradhan@iucaa.ernet.in, acpradhan@yahoo.com}}\\
\vspace{5mm}
\normalsize{$^{\dag}$, $^{\ddag}$\it{ Department of Mathematics, Post-graduate College, 
Ghazipur-233 001, India \\
E-mail : vandana\_rai005@yahoo.co.in}}\\
\end{center}
\vspace{10mm}
\begin{abstract} 
A new class of exact solutions of Einstein's field equations with bulk viscous 
fluid for an LRS Bianchi type-I spacetime is obtained by using deceleration 
parameter as variable. The value of Hubble's constant $H_{0}$ is found to be less than 
one for non-flat model and is equal to $1.5$ for flat model which are of the physical 
interest. Some physical and geometric properties of the models are also discussed.
\end{abstract}
\smallskip
\n PACS No. : 04.20.-q, 98.80.-k \\
\n Keywords : Cosmology, LRS Bianchi type-I universe, bulk viscous models, deceleration parameter.
\section{Introduction}
The Bianchi cosmologies play an important role in theoretical cosmology and 
have been much studied since the 1960s. A Bianchi cosmology represents a spatially 
homogeneous universe, since by definition the spacetime admits a three-parameter 
group of isometries whose orbits are space-like hyper-surfaces. These models can be
used to analyze aspects of the physical Universe which pertain to or which may be 
affected by anisotropy in the rate of expansion, for example , the cosmic microwave 
background radiation, nucleosynthesis in the early universe, and the question of the 
isotropization of the universe itself \cite{ref1}. For simplification and description 
of the large scale behaviour of the actual universe, locally rotationally symmetric 
[henceforth referred as LRS] Bianchi I spacetime have widely studied 
{\cite{ref2}$-$\cite{ref6}}. When the Bianchi I spacetime expands equally in two 
spatial directions it is called locally rotationally symmetric. These kinds of models 
are interesting because Lidsey \cite{ref7} showed that they are equivalent to a flat 
(FRW) universe with a self-interesting scalar field and a free massless scalar field, 
but produced no explicit example. Some explicit solutions were pointed out in references 
\cite{ref8,ref9}.
\newline
\par
The Einstein's field equations are coupled system of high non-linear differential 
equations and we seek physical solutions to the field equations for their applications 
in cosmology and astrophysics. In order to solve the field equations we normally assume 
a form for the matter content or that spacetime admits killing vector symmetries \cite{ref10}. 
Solutions to the field equations may also be generated by applying a law of variation for 
Hubble's parameter which was proposed by Berman \cite{ref11}. In simple cases the Hubble 
law yields a constant value of deceleration parameter. It is worth observing that most of 
the well-known models of Einstein's theory and Brans-Deke theory with curvature parameter 
$k = 0$, including inflationary models, are models with constant deceleration parameter. 
In earlier literature cosmological models with a constant deceleration parameter have been 
studied by Berman \cite{ref11}, Berman and Gomide \cite{ref12}, Johri and Desikan \cite{ref13}, 
Singh and Desikan \cite{ref14}, Maharaj and Naidoo \cite{ref15}, Pradhan and {\it et al.} \cite{ref16} 
and others. But redshift magnitude test has had a chequered history. During the 1960s and the 1970s, 
it was used to draw very categorical conclusions. The deceleration parameter $q_{0}$ was then 
claimed to lie between $0$ and $1$ and thus it was claimed that the universe is decelerating. 
Today's situation, we feel, is hardly different. Observations \cite{ref17,ref18} of Type Ia 
Supernovae (SNe) allow to probe the expansion history of the universe. The main conclusion of 
these observations is that the expansion of the universe is accelerating. So we can consider 
the cosmological models with variable deceleration parameter. The readers are advised to see the 
papers by Vishwakarma and Narlikar \cite{ref19} and Virey {\it et al.} \cite{ref20} and references 
therein for a review on the determination of the deceleration parameter from Supernovae data.    
\newline
\par
Most cosmological models assume that the matter in the universe can be described 
by `dust'(a pressure-less distribution) or at best a perfect fluid. The role of the bulk 
viscosity in the cosmic evolution, especially as its early stages, seems to be significant. 
The general criterion for bulk viscosity was given by Israel and Vardalas \cite{ref21}, 
Klimek \cite{ref22} and Weinberge \cite{ref23}). For example, the existence of the bulk 
viscosity is equivalent to slow process of restoring equilibrium states 
(Landau and Lifshitz \cite{ref24}). The presently observed high entropy per baryon ratio 
in the universe can be explained by involving some kind of dissipative mechanism (e. g., 
bulk viscosity). Bulk viscosity is associated with the GUT phase transition and string creation. 
Thus, we should consider the presence of a material distribution other than a perfect 
fluid to have realistic cosmological models (see Gr\o n \cite{ref25}) for a review on 
cosmological models with bulk viscosity). The model studied by Murphy \cite{ref26} possessed 
an interesting feature in that the big bang type of singularity of infinite spacetime 
curvature does not occur to be a finite past. However, the relationship assumed by Murphy 
between the viscosity coefficient and the matter density is not acceptable at large 
density. The effect of bulk viscosity on the cosmological evolution has been 
investigated by a number of authors in the framework of general theory of 
relativity (Padmanabhan and Chitre \cite{ref27}, Pavon \cite{ref28}, Johri and Sudarshan
\cite{ref29}, Maartens \cite{ref30}, Zimdahl \cite{ref31}, Santos {\it et al.} \cite{ref32},
Pradhan, Sarayakar and Beesham \cite{ref33}, Kalyani and Singh \cite{ref34},
Singh, Beesham and Mbokazi \cite{ref35}, Pradhan {\it et al.} \cite{ref36}), Singh {\it et al.} 
\cite{ref37}), Bali and Pradhan \cite{ref38}). This motivates to study cosmological bulk viscous 
fluid model.  
\newline
\par
Recently Paul \cite{ref39} have investigated LRS Bianchi type-I cosmological models with 
a variable deceleration parameter. In this paper, we propose to find LRS Bianchi type-I 
cosmological models in presence of a bulk viscous fluid and we will generalize the 
solutions \cite{ref39}.\\ 
\section{The Metric and Field  Equations}
We consider the LRS Bianchi type-I  metric in the form \cite{ref5}
\begin{equation}
\label{eq1}
ds^{2} = dt^{2} - A^{2} dx^{2} - B^{2}(dy^{2} + dz^{2}),
\end{equation}
where A and B are functions of $x$ and $t$.
The stress energy-tensor in the presence of bulk stress has the form 
\begin{equation}
\label{eq2}
T_{ij} = (\rho + \bar{p})u_{i}u_{j} - \bar{p} g_{ij},  
\end{equation}
where
\begin{equation}
\label{eq3}
\bar{p} = p - \xi u^{i}_{;i}.
\end{equation}
Here $\rho$, $p$, $\bar{p}$ and $\xi$ are the energy density,
thermodynamical pressure, effective pressure and  bulk viscous 
coefficient respectively and $u_{i}$ is the four velocity  vector satisfying 
the relations
\begin{equation}
\label{eq4}
u_{i}u^{i} = 1,
\end{equation}
The Einstein's field equations (in gravitational units $c = 1$, $G = 1$) read as
\begin{equation}
\label{eq5}
R_{ij} - \frac{1}{2} R g_{ij} = -8\pi T_{ij},
\end{equation}
where $R_{ij}$ is the Ricci tensor; $R$ = $g^{ij} R_{ij}$ is the
Ricci scalar. The Einstein's field equations (\ref{eq5}) for the line element (\ref{eq1})
has been set up as
\begin{equation}
\label{eq6}
\frac{2\ddot{B}}{B} + \frac{\dot{B}^{2}}{B^{2}} - \frac{{B^{\prime}}^{2}}{A^{2}B^{2}} = 
- 8\pi \bar{p},
\end{equation}
\begin{equation}
\label{eq7}
{\dot{B}}^{\prime} - \frac{B^{\prime}\dot{A}}{A} = 0,
\end{equation}
\begin{equation}
\label{eq8}
\frac{\ddot{A}}{A} + \frac{\ddot{B}}{B} + \frac{\dot{A}\dot{B}}{AB} - \frac{B^{\prime \prime}}
{A^{2}B} + \frac{A^{\prime}B^{\prime}}{A^{3}B} = - 8\pi \bar{p},
\end{equation}
\begin{equation}
\label{eq9}
\frac{2 B^{\prime \prime}}{A^{2}B} - \frac{2 A^{\prime}B^{\prime}}{A^{3}B} + 
\frac{{B^{\prime}}^{2}}{A^{2}B^{2}} - \frac{2\dot{A}\dot{B}}{AB} - \frac{\dot{B}^{2}}{B^{2}}
= - 8\pi \rho. 
\end{equation}
The energy conservation equation yields
\begin{equation}
\label{eq10}
\dot{\rho} + (\bar{p} + \rho)\left(\frac{\dot{A}}{A} + \frac{2\dot{B}}{B}\right) = 0,
\end{equation}
where dots and primes indicate partial differentiation with respect to $t$ and $x$ 
respectively. 
\section{Solution of the Field Equations}
Equation (\ref{eq7}), after integration, yields
\begin{equation}
\label{eq11}
A = \frac{B^{\prime}}{\ell},
\end{equation}
where $\ell$ is an arbitrary function of $x$. Equations (\ref{eq6}) and (\ref{eq8}), 
with the use of Eq. (\ref{eq11}), reduces to 
\begin{equation}
\label{eq12}
\frac{B}{B^{\prime}}\frac{d}{dx}\left(\frac{\ddot{B}}{B}\right) + \frac{\dot{B}}{B^{\prime}}
\frac{d}{dt}\left(\frac{B^{\prime}}{B}\right) + \frac{\ell^{2}}{B^{2}}\left(1 - \frac{B}{B^{\prime}}
\frac{\ell^{\prime}}{\ell}\right) = 0.
\end{equation}
Since $A$ and $B$ are separable functions of $x$ and $t$, so, $\frac{B^{\prime}}{B}$ is a 
function of $x$ alone. Hence, after integrating Eq. (\ref{eq12}) gives
\begin{equation}
\label{eq13}
B = \ell S(t),
\end{equation}
where $S$ is a scale factor which is an arbitrary function of $t$. Thus from Eqs. (\ref{eq11}) 
and (\ref{eq13}), we have
\begin{equation}
\label{eq14}
A = \frac{\ell^{\prime}}{\ell} S.
\end{equation}
Now the metric (\ref{eq1}) is reduced to the form
\begin{equation}
\label{eq15}
ds^{2} = dt^{2} - S^{2}\left[dX^{2} + e^{2X}(dy^{2} + dz^{2})\right],
\end{equation}
where $X = ln~{\ell}$. The mass-density, effective pressure and Ricci scalar are obtained as
\begin{equation}
\label{eq16}
8\pi \rho = \frac{3}{S^{2}}\left[{\dot{S}}^2 - 1\right],
\end{equation}
\begin{equation}
\label{eq17}
8\pi \bar{p} = \frac{1}{S^{2}}\left[1 - {\dot{S}}^2 - 2 S \ddot{S}\right],
\end{equation}
\begin{equation}
\label{eq18}
R = \frac{6}{S^{2}}\left[S \ddot{S} + {\dot{S}}^2 - 1\right].
\end{equation}
The function $S(t)$ remains undetermined. To obtain its explicit dependence on $t$, 
one may have to introduce additional assumption. To achieve this, we assume the 
deceleration parameter to be variable, i.e.
\begin{equation}
\label{eq19}
q = - \frac{S\ddot{S}}{{\dot{S}}^2} = - \left(\frac{\dot{H} + H^{2}}{H^{2}}\right) = b (variable),
\end{equation} 
where $H = \frac{\dot{S}}{S}$ is the Hubble parameter. The above equation may be rewritten as
\begin{equation}
\label{eq20}
\frac{\ddot{S}}{S}+ b \frac{{\dot{S}}^2}{S^{2}} = 0.
\end{equation} 
\section{Solution for $b = -\frac{a S}{{\dot{S}}^2}$, where $a$ is constant} 
In this case, on integrating, Eq. (\ref{eq20}) gives the exact solution
\begin{equation}
\label{eq21}
S = \frac{1}{2} at^{2} + kt + d,
\end{equation} 
where $k$ and $d$ are constants of integration. Here we consider the two following cases: 
\subsection{Non-flat models} 
The mass-density, pressure and Ricci scalar are given by 
\begin{equation}
\label{eq22}
8\pi \rho = \frac{3}{S^{2}}\left[(k + at)^{2} - 1\right],
\end{equation}
\begin{equation}
\label{eq23}
8\pi (p - \xi \theta) = - \frac{1}{S^{2}}\left[(k + at)^{2} - 1 + 2a(\frac{1}{2}at^{2} + kt + d) \right],
\end{equation}
\begin{equation}
\label{eq24}
R = \frac{6}{S^{2}}\left[(k + at)^{2} - 1 + a(\frac{1}{2}at^{2} + kt + d)\right].
\end{equation}
Here $\xi$, in general, is a function of time. The expression for kinematical parameter 
expansion $\theta$ is given by
\begin{equation}
\label{eq25}  
\theta = \frac{3(k + at)}{S}.
\end{equation}
Thus, given $\xi(t)$ we can solve the equations. In most of the investigations involving bulk 
viscosity is assumed to be a simple power function of the energy density \cite{ref28}$-$\cite{ref30}
\begin{equation}
\label{eq26} 
\xi(t) = \xi_{0} \rho^{n},
\end{equation}
where $\xi_{0}$ and $n$ are constants. For small density, $n$ may even be equal to unity 
as used in Murphy's work for simplicity \cite{ref7}. If $n = 1$, Eq. (\ref{eq26}) may 
correspond to a radiative fluid \cite{ref23}. However, more realistic models \cite{ref32} 
are based on $n$ lying in the regime $0 \leq n \leq \frac{1}{2}$. 

On thermodynamical grounds, in conventional physics $\xi$ has to be positive; this being a 
consequence of the positive entropy change in irreversible processes. For simplicity and 
realistic models of physical importance, we consider the following two subcases $(n = 0, 1)$: 
\subsubsection{Model I: ~ ~ solution for  $\xi = \xi_{0}$} 
When $n = 0$, Eq. (\ref{eq26}) reduces to $\xi = \xi_{0}$ = constant. Hence in this case 
Eq. (\ref{eq23}), with the use of (\ref{eq25}), leads to
\begin{equation}
\label{eq27} 
8\pi p = \frac{24\pi \xi_{0}}{S}(k + at) - \frac{1}{S^{2}}\left[(k + at)^{2} - 1 + 2a(\frac{1}{2}at^{2} 
+ kt + d)\right].
\end{equation}
\subsubsection{Model II: ~ ~ solution for $\xi = \xi_{0}\rho$} 
When $n = 1$, Eq. (\ref{eq26}) reduces to $\xi = \xi_{0}\rho$. Hence in this case 
Eq. (\ref{eq23}), with the use of (\ref{eq25}), reduces to
\begin{equation}
\label{eq28} 
8\pi p = \frac{\{(k + at)^{2} - 1\}}{S^{3}}\left[9\xi_{0}(k + at) - S\right] - \frac{2a(\frac{1}{2}at^{2} 
+ kt + d)}{S^{2}}.
\end{equation}
\subsubsection{Physical behaviour of the models} 
The effect of the bulk viscosity is to produce a change in perfect fluid and hence 
exhibit essential influence on the character of the solution. We also observe here 
that Murphy's condition \cite{ref7} about the absence of a big bang type singularity 
in the finite past in models with bulk viscous fluid, in general, not true. \\
From Eqs. (\ref{eq22}) and (\ref{eq24}), it is observed that 
\begin{equation}
\label{eq29} 
\rho > 0 ~ ~  and ~ ~   R > 0 ~ ~ for ~ ~  t > \frac{1 - k}{a},
\end{equation}
where $k < 1$. From Eqs. (\ref{eq22}), (\ref{eq24}) and (\ref{eq25}), it is also 
observed that $\rho$, $R$ and $\theta$ decrease as $t$ increases. 

We find that shear $\sigma = 0$ in the models. Hence $\sigma/\theta = 0$, which shows that 
the models are isotropic. From Hubble's parameter equation, $H = \frac{\dot{S}}{S}$, we have 
an epoch time $t_{0}$ given by
\begin{equation}
\label{eq30} 
t_{0} = \frac{1}{H_{0}} - \frac{k}{a} + \frac{\sqrt{a^{2} + {H_{0}}^2(k^{2} - 2a d)}}{a H_{0}},
\end{equation}  
which gives that
\begin{equation}
\label{eq31} 
a t_{0} + k = \frac{a}{H_{0}} - 1 + \frac{\sqrt{a^{2} + {H_{0}}^2(k^{2} - 2a d)}}{H_{0}} > 0.
\end{equation} 
From above equation we conclude that 
$$
\frac{a}{H_{0}} - 1 > 0,
$$
which reduces to
\begin{equation}
\label{eq32} 
H_{0} <  a.
\end{equation}
From Eq. (\ref{eq30}), we observe that $t_{0} > 0$ for $H_{0} < \frac{a}{k}$ and $k^{2} > 2ad$ 
i.e. 
$$H_{0} < \frac{k}{2d}.$$ 
From Eq. (\ref{eq24}), $R > 0$ implies that 
\begin{equation}
\label{eq33} 
(k + at)^{2} > [\sqrt{1 - a(\frac{1}{2}at^{2} + kt + d)}]^{2}.
\end{equation}
It is evident from Eqs. (\ref{eq29}) and (\ref{eq33}) that $a < 1$ and hence from (\ref{eq32}) 
we obtain
\begin{equation}
\label{eq34} 
H_{0} < 1.
\end{equation}
The models, in general, represent expanding, non-shearing and isotropic universe. The models 
in the presence of bulk viscosity start expanding with a big bang at $t = 0$ when $d = 0$ and 
the expansion in the models decreases as time increases and the expansion stops at $t = \infty$ 
and $t = -\frac{k}{a}$. \\
\subsection{Flat models} 
For flat model Ricci scalar, $R = 0$ and hence in this case
\begin{equation}
\label{eq35} 
(k + at)^{2} = [1 - a(\frac{1}{2}at^{2} + kt + d)].
\end{equation}
The mass-density and pressure are given by 
\begin{equation}
\label{eq36}
8\pi \rho = \frac{3}{S^{2}}\left[(k + at)^{2} - 1\right],
\end{equation}
\begin{equation}
\label{eq37}
8\pi (p - \xi \theta) = \frac{1}{S^{2}}\left[(k + at)^{2} - 1\right].
\end{equation}
The expansion $\theta$ is given
\begin{equation}
\label{eq38}
\theta = \frac{3\sqrt{[1 - a(\frac{1}{2}at^{2} + kt + d)]}}{S}.
\end{equation}
Here we again consider two subcases: \\
\subsubsection{Model I: ~ ~ solution for  $\xi = \xi_{0}$} 
When $n = 0$, Eq. (\ref{eq26}) reduces to $\xi = \xi_{0}$ = constant. Hence in this case 
Eq. (\ref{eq37}), with the use of (\ref{eq38}), leads to
\begin{equation}
\label{eq39} 
8\pi p = \frac{24\pi \xi_{0}}{S}\sqrt{[1 - a(\frac{1}{2}at^{2} + kt + d)]}
 + \frac{1}{S^{2}}\left[(k + at)^{2} - 1\right].
\end{equation}
\subsubsection{Model II: ~ ~ solution for $\xi = \xi_{0}\rho$} 
When $n = 1$, Eq. (\ref{eq26}) reduces to $\xi = \xi_{0}\rho$. Hence in this case 
Eq. (\ref{eq37}), with the use of (\ref{eq38}), reduces to
\begin{equation}
\label{eq40} 
8\pi p = \frac{\{(k + at)^{2} - 1\}}{S^{3}}\left[9\xi_{0}\sqrt{[1 - a(\frac{1}{2}at^{2} + kt + d)]} 
+ S\right].
\end{equation}
The models, in general, represent expanding, non-shearing and isotropic universes. The models 
in the presence of bulk viscosity start expanding with a big bang at $t = 0$ when $d = 0$ and 
the expansion in the models decreases as time increases and the expansion stops at $t = \infty$ 
and $a = 1$.
\section{Solution for $b = -\frac{a t S}{{\dot{S}}^2}$, where $a$ is constant} 
In this case, on integrating, Eq. (\ref{eq20}) gives the exact solution
\begin{equation}
\label{eq41}
S = \frac{1}{6} at^{3} + kt + d,
\end{equation} 
where $k$ and $d$ are constants of integration. \\

 Here we consider the two following cases: 
\subsection{Non-flat models} 
The mass-density, pressure and Ricci scalar are given by 
\begin{equation}
\label{eq42}
8\pi \rho = \frac{3}{S^{2}}\left[\left(k + \frac{1}{2}at^{2}\right)^{2} - 1\right],
\end{equation}
\begin{equation}
\label{eq43}
8\pi(p - \xi \theta) = -\frac{1}{S^{2}}\left[\left(k + \frac{1}{2}at^{2}\right)^{2} - 1 + 
2at(\frac{1}{6}at^{3} + kt + d)\right],
\end{equation}
\begin{equation}
\label{eq44}
R = \frac{6}{S^{2}}\left[\left(k + \frac{1}{2}at^{2}\right)^{2} - 1 + at(\frac{1}{6}at^{3} + kt + d) \right].
\end{equation}
The expression for kinematical parameter expansion $\theta$ is given by
\begin{equation}
\label{eq45}
\theta = \frac{3}{S}\left(k + \frac{1}{2}at^{2}\right),
\end{equation} 
where $S$ is given by (\ref{eq41}). \\

For simplicity and realistic models of physical importance, we consider the following two 
subcases: \\
\subsubsection{Model I: ~ ~ solution for  $\xi = \xi_{0}$}
When $n = 0$, Eq. (\ref{eq26}) reduces to $\xi = \xi_{0}$ = constant. Hence in this case 
Eq. (\ref{eq43}), with the use of (\ref{eq45}), leads to
\begin{equation}
\label{eq46} 
8\pi p = \frac{24\pi \xi_{0}}{S}(k + \frac{1}{2}at^{2}) - \frac{1}{S^{2}}\left[(k + \frac{1}
{2}at^{2})^{2} - 1 + 2at(\frac{1}{6}at^{3} + kt + d) \right].
\end{equation}
\subsubsection{Model II: ~ ~ solution for $\xi = \xi_{0}\rho$} 
When $n = 1$, Eq. (\ref{eq26}) reduces to $\xi = \xi_{0}\rho$. Hence in this case 
Eq. (\ref{eq43}), with the use of (\ref{eq45}), reduces to
\begin{equation}
\label{eq47} 
8\pi p = \frac{\{(k + \frac{1}{2}at^{2})^{2} - 1\}}{S^{3}}\left[9\xi_{0}(k + \frac{1}{2}at^{2}) 
- S\right] - \frac{2a(\frac{1}{6}at^{3} + kt + d)}{S^{2}}.
\end{equation}
\subsubsection{Physical behaviour of the models}
From Eqs. (\ref{eq42}) and (\ref{eq44}), it is observed that $\rho > 0$ and $R > 0$ for 
$t > \sqrt{(2/a)(1 - k)}$, where $k < 1$. From (\ref{eq45}), we observe that $\theta$ decreases 
as $t$ increases. \\

We find that shear $\sigma = 0$ in the models. Hence $\sigma/\theta = 0$ which shows that the models 
are isotropic. At any intermediate time $t = \sqrt{(2/a)(2 - k)}$, $R > 0$ implies that  
\begin{equation}
\label{eq48}
3 + \frac{4}{3}(2- k)(1 + k) + d\sqrt{2a(2 - k)} > 0,
\end{equation} 
where $k < 2$. From Eq. (\ref{eq48}), it is evident that
\begin{equation}
\label{eq49}
d\sqrt{2a(2 - k)} >0.
\end{equation}
From Hubble's parameter, $H = \frac{\dot{S}}{S}$, we obtain a cubic equation in $t_{0}$ 
\begin{equation}
\label{eq50}
{t_{0}}^3 - \frac{3{t_{0}}^2}{H_{0}} + \frac{6k t_{0}}{a} - \frac{6}{aH_{0}}\left(k - dH_{0}
\right) = 0.
\end{equation}
Solving Eq. (\ref{eq50}) we obtain
\begin{equation}
\label{eq51}
t_{0} = \frac{1}{H_{0}}, ~ ~ a = 2k{H_{0}}^2 ~ ~ and ~ ~ d = \frac{a}{3{H_{0}}^3}, 
\end{equation}
where $t_{0}$ is an epoch time. Thus from Eq. (\ref{eq49})
\begin{equation}
\label{eq52}
4{H_{0}}^{2} k(2 - k) > 0.
\end{equation}
Hence Eq. (\ref{eq52}), for $k<2$, implies that
\begin{equation}
\label{eq53}
H_{0} > 0.
\end{equation}
The models, in general, represent expanding, non-shearing and isotropic universe. The models 
in the presence of bulk viscosity start expanding with a big bang at $t = 0$ when $d = 0$ and 
the expansion in the models decreases as time increases and the expansion stops at $t = \infty$ 
and $t^2 = -\frac{2k}{a}$.
\subsection{Flat model} 
For flat model Ricci scalar, $R = 0$ and hence in this case
\begin{equation}
\label{eq54} 
(k + \frac{1}{2}at^{2})^{2} = [1 - at(\frac{1}{6}at^{3} + kt + d)].
\end{equation}
The mass-density and pressure are given by 
\begin{equation}
\label{eq55}
8\pi \rho = \frac{3}{S^{2}}\left[(k + \frac{1}{2}at^{2})^{2} - 1\right],
\end{equation}
\begin{equation}
\label{eq56}
8\pi (p - \xi \theta) = \frac{1}{S^{2}}\left[(k + \frac{1}{2}at^{2})^{2} - 1\right].
\end{equation}
The expansion $\theta$ is given
\begin{equation}
\label{eq57}
\theta = \frac{3\sqrt{[1 - at(\frac{1}{6}at^{3} + kt + d)]}}{S}.
\end{equation}
Here we again consider two subcases: \\
\subsubsection{Model I: ~ ~ solution for  $\xi = \xi_{0}$} 
When $n = 0$, Eq. (\ref{eq26}) reduces to $\xi = \xi_{0}$ = constant. Hence in this case 
Eq. (\ref{eq56}), with the use of (\ref{eq57}), leads to
\begin{equation}
\label{eq58} 
8\pi p = \frac{24\pi \xi_{0}}{S}\sqrt{[1 - at(\frac{1}{6}at^{3} + kt + d)]}
 + \frac{1}{S^{2}}\left[(k + \frac{1}{2}at^{2})^{2} - 1\right].
\end{equation}
\subsubsection{Model II: ~ ~ solution for $\xi = \xi_{0}\rho$} 
When $n = 1$, Eq. (\ref{eq26}) reduces to $\xi = \xi_{0}\rho$. Hence in this case 
Eq. (\ref{eq56}), with the use of (\ref{eq57}), reduces to
\begin{equation}
\label{eq59} 
8\pi p = \frac{\{(k + \frac{1}{2}at^{2})^{2} - 1\}}{S^{3}}\left[9\xi_{0}
\sqrt{[1 - at(\frac{1}{6}at^{3} + kt + d)]} + S\right].
\end{equation}
The models, in general, represent expanding, non-shearing and isotropic universe. The 
energy conditions $\rho > 0$ and $p > 0$ are satisfied when $\sqrt{(2/a)(1 - k)}$.
The models in the presence of bulk viscosity start expanding with a big bang at $t = 0$ 
when $d = 0$ and the expansion in the models decreases as time increases and the expansion 
stops at $t = \infty$ and $t^2 = -\frac{2k}{a}$.
\section{Solution for $b = -\frac{K S}{{\dot{S}}^3}$, where $K$ is constant} 
In this case, on integrating, Eq. (\ref{eq20}) gives the exact solution
\begin{equation}
\label{eq60}
S = \beta + \frac{(\alpha + 2 Kt)^{3/2}}{3K},
\end{equation} 
where $\alpha$ and $\beta$ are constants of integration. \\

Here we consider the two following cases: 
\subsection{Non-flat models} 
The mass-density, pressure and Ricci scalar are given by 
\begin{equation}
\label{eq61}
8\pi \rho = \frac{3}{S^{2}}\left[(\alpha + 2Kt) - 1 \right],
\end{equation}
\begin{equation}
\label{eq62}
8\pi (p - \xi \theta) = \frac{1}{S^{2}}\left[1 - (\alpha + 2Kt) - \frac{2K}{(\alpha + 2Kt)
^{1/2}}\{\beta + \frac{(\alpha + 2Kt)^{\frac{3}{2}}}{3K}\}\right],
\end{equation}
\begin{equation}
\label{eq63}
R = \frac{6}{S^{2}}\left[(\alpha + 2Kt) - 1 +  \frac{K}{(\alpha + 2Kt)^{1/2}}  
\{\beta + \frac{(\alpha + 2Kt)^{\frac{3}{2}}}{3K}\} \right].
\end{equation}
The expansion $\theta$ is calculated as
\begin{equation}
\label{eq64}
\theta = \frac{3}{S}(\alpha + 2Kt)^{1/2}.
\end{equation}
For simplicity and realistic models of physical importance, we consider the following two 
subcases: \\
\subsubsection{Model I: ~ ~ solution for  $\xi = \xi_{0}$} 
When $n = 0$, Eq. (\ref{eq26}) reduces to $\xi = \xi_{0}$ = constant. Hence in this case 
Eq. (\ref{eq62}), with the use of (\ref{eq66}), leads to
\begin{equation}
\label{eq65} 
8\pi p = \frac{24\pi \xi_{0}}{S} (\alpha + 2Kt)^{1/2} - \frac{1}{S^{2}}\left[\alpha + 2Kt - 1 
+ \frac{2K}{(\alpha + 2Kt)^{1/2}}\{\beta + \frac{(\alpha + 2Kt)^{\frac{3}{2}}}{3K}\} \right].
\end{equation}
\subsubsection{Model II: ~ ~ solution for $\xi = \xi_{0}\rho$} 
When $n = 1$, Eq. (\ref{eq26}) reduces to $\xi = \xi_{0}\rho$. Hence in this case 
Eq. (\ref{eq62}), with the use of (\ref{eq64}), reduces to
\begin{equation}
\label{eq66} 
8\pi p = \frac{(\alpha + 2Kt -1)}{S^{3}}\left[9(\alpha + 2Kt)^{1/2} - S\right] - 
\frac{2K}{S^{2}(\alpha + 2Kt)^{1/2}}\{\beta + \frac{(\alpha + 2Kt)^{\frac{3}{2}}}{3K}\} .
\end{equation}
\subsubsection{Physical behaviour of the models} 
From Eq. (\ref{eq61}), it is observed that $\rho > 0$ for $t > (1 - \alpha)/2K$, where 
$\alpha < 1$. From (\ref{eq64}), we observe that $\theta$ decreases as $t$ increases. \\

We find that shear $\sigma = 0$ in the models. Hence $\sigma/\theta = 0$ which shows that the models 
are isotropic in nature. \\
From the Hubble's parameter, $H = \frac{\dot{S}}{S}$, we obtain
\begin{equation}
\label{eq67} 
{\chi_{0}}^3 - 3K\left(\frac{\chi_{0}}{H_{0}} - \beta\right) = 0,
\end{equation}
where $\chi_{0} = \sqrt{\alpha + 2Kt_{0}}$ and $t_{0}$ is an epoch time. For the solution of 
Eq. (\ref{eq67}) let us assume $\chi_{0} = \beta$ and hence we have 
\begin{equation}
\label{eq68} 
\beta^{2} = 3K\left(\frac{1}{H_{0}} - 1\right).
\end{equation}
From Eq. (\ref{eq63}), $R > 0$ implies that 
\begin{equation}
\label{eq69} 
4\chi^{3} - 3\chi + 3\beta K > 0,
\end{equation}
where $\chi = \sqrt{\alpha + 2Kt}$. \\
Solving Eq. (\ref{eq69}) at any intermediate time, $t = (3 - \alpha)/2K$, where $\alpha < 3$, 
we obtain
\begin{equation}
\label{eq70} 
3 + K > 0.
\end{equation}
From $\chi_{0} = \sqrt{\alpha + 2Kt_{0}} = \beta$, we have
$$\alpha + 2K\frac{3 - \alpha}{2K} = \beta^{2},$$
which reduces to
\begin{equation}
\label{eq71} 
\beta^{2} = 3. 
\end{equation}
From Eqs. (\ref{eq68}) and (\ref{eq71}), we obtain
\begin{equation}
\label{eq72} 
K = \frac{1}{\frac{1}{H_{0}} - 1}.
\end{equation}
From Eqs. (\ref{eq70}) and (\ref{eq72}), we have
\begin{equation}
\label{eq73} 
H_{0} < 1.5.
\end{equation}
From (\ref{eq72}) in order that $K > 0$, $H_{0} < 1$. Hence the value of Hubble's constant is 
less than unity.
\subsection {Flat model} 
For flat model, Ricci scalar $R = 0$ gives at any intermediate time, $t = (3 - \alpha_{1})/2k_{1}$,
where $\alpha_{1} < 3$,
\begin{equation}
\label{eq74} 
3 + k_{1} = 0.
\end{equation}
The above equation with the use of Eq. (\ref{eq72}) gives
\begin{equation}
\label{eq75} 
H_{0} = 1.5
\end{equation}
The mass-density and pressure are given by
\begin{equation}
\label{eq76} 
8\pi \rho = \frac{3}{S^{2}}\left[(\alpha_{1} + 2k_{1}t) - 1\right], 
\end{equation}
\begin{equation}
\label{eq77} 
8\pi (p - \xi \theta) = \frac{1}{S^{2}}\left[(\alpha_{1} + 2k_{1}t) - 1\right], 
\end{equation}
where
\begin{equation}
\label{eq78} 
S = \beta_{1} + \frac{(\alpha_{1} + 2k_{1}t)^{3/2}}{3k_{1}}, 
\end{equation}
and $k_{1}$, $\alpha_{1}$ and $\beta_{1}$ are constants. In non-flat model, Eq. (\ref{eq71}) 
gives $\beta_{1} = \sqrt{3}$ and from (\ref{eq75}) using $H_{0} = 1.5$, we have from (\ref{eq72}), 
$k_{1} = - 3$. Accordingly $S$ is positive for
$$ t > \left(\alpha_{1} - (3\beta_{1}k_{2})^{2/3}\right)/2k_{1},$$
where $k_{2} = 3$ such that $(3\beta_{1}k_{2})^{2/3} < \alpha_{1} < 3$.
\section{Conclusions}
In this paper we have described a new class of LRS Bianchi type I cosmological models 
with a bulk viscous fluid as the source of matter by applying variable deceleration 
parameter. Generally, the models are expanding, non-shearing and isotropic in nature. 
The models in the presence of bulk viscosity start expanding with a big bang at $t = 0$ 
when $d = 0$ and the expansion in the models decreases as time increases and the expansion 
stops at $t = \infty$ and $t = -\frac{k}{a}$. The study of the results of the three deceleration 
parameter models of the universe showed that the Hubble's constant is less than $1.00$ for 
the non-flat model, and it is $1.5$ for the flat model. These mathematical results are of 
the physical interest. The scale factor is not obtained to be linearly related to time as 
the case with supernova cosmology model \cite{ref40}. If $\xi = 0$ is set in the solutions
obtained in this paper, we get the solutions obtained by Paul \cite{ref39}. But in his 
paper \cite{ref39}, there are errors in the the beginning equations (\ref{eq15}) and 
(\ref{eq16}) which propagate through out the paper and affect all the results.    

The coefficient of bulk viscosity is taken to be a power function of mass density.
The effect of the bulk viscosity is to produce a change in perfect fluid and hence 
exhibit essential influence on the character of the solution. Murphy \cite{ref26}  has 
studied perfect fluid cosmological models with bulk viscosity and obtained that the 
big bang singularity may be avoided in the finite past. We also observe here 
that Murphy's condition \cite{ref7} about the absence of a big bang type singularity 
in the finite past in models with bulk viscous fluid, in general, not true. 
\section*{Acknowledgements} 
A. Pradhan thanks to Prof. T. R. Seshadri, I.R.C., Delhi University, 
India for providing  facility where part of work was carried out. The authors 
would like to thank A. K. Yadav for fruitful discussions. The authors also thank to the 
referees for the fruitful comments which improved the paper in this form.\\
\newline
\newline

\end{document}